\documentclass[preprint,aps,amsmath,amssymb,showpacs]{revtex4-1}
\usepackage{graphicx}
\usepackage{xcolor}

\begin{document}
\title{Polymers with nearest- and next nearest-neighbor interactions on the Husimi lattice}
\author{Tiago J. Oliveira\footnote{On leave at Ames Laboratory - USDOE and Department of Physics \& Astronomy, Iowa State University, Ames, Iowa 50011, United States}}
\email{tiago@ufv.br}
\affiliation{Departamento de F\'isica, Universidade Federal de Vi\c cosa, 36570-900, Vi\c cosa, MG, Brazil}
\date{\today}

\begin{abstract}
The exact grand-canonical solution of a generalized interacting self-avoid walk (ISAW) model, placed on a Husimi lattice built with squares, is presented. In this model, beyond the traditional interaction $\omega_1=e^{\epsilon_1/k_B T}$ between (nonconsecutive) monomers on nearest-neighbor (NN) sites, an additional energy $\epsilon_2$ is associated to next-NN (NNN) monomers. Three definitions of NNN sites/interactions are considered, where each monomer can have, effectively, at most 2, 4 or 6 NNN monomers on the Husimi lattice. The phase diagrams found in all cases have (qualitatively) the same thermodynamic properties: a non-polymerized (NP) and a polymerized (P) phase separated by a critical and a coexistence surface that meet at a tricritical ($\theta$-) line. This $\theta$-line is found even when one of the interactions is repulsive, existing for $\omega_1$ in the range $[0,\infty)$, i. e., for $\epsilon_1/k_B T$ in the range $[-\infty,\infty)$. Counterintuitively, a $\theta$-point exists even for an infinite repulsion between NN monomers ($\omega_1=0$), being associated to a coil-``soft globule'' transition. In the limit of an infinite repulsive force between NNN monomers, however, the coil-globule transition disappears and only a NP-P continuous transition is observed. This particular case, with $\omega_2=0$, is also solved exactly on the square lattice, using a transfer matrix calculation, where a discontinuous NP-P transition is found. For attractive and repulsive forces between NN and NNN monomers, respectively, the model becomes quite similar to the semiflexible-ISAW one, whose crystalline phase is not observed here, as a consequence of the frustration due to competing NN and NNN forces. The mapping of the phase diagrams in canonical ones is discussed and compared with recent results from Monte Carlo simulations.
\end{abstract}

\maketitle

\section{Introduction}
\label{intro}

A polymer in solution, usually, may exist in three different conformations depending on temperature $T$ or solvent quality: \textit{i) extended}: \textit{coil}-like chains (high $T$ and/or good solvents); \textit{ii) collapsed}: the chains have \textit{globule}-like shapes (low $T$ and/or poor solvents); and \textit{iii) $\theta$}: this point marks the (continuous/tricritical \cite{DeGennes1,DeGennes2}) transition between the coil and globule phases (occurring at $T_{\theta}$ and in a ``$\theta$-solvent'') \cite{Flory1,Flory2}. The differences among these phases can be characterized, for example, through the metric exponent $\nu$ - from the scaling of the gyration radio $R_g$ with the number $N$ of monomers, $R_g \sim N^{\nu}$ -, being $\nu_{coil} > \nu_{\theta} > \nu_{globule}$.

For linear polymers, the coil phase can be modeled by self-avoiding walks (SAWs), where the excluded volume is the only interaction present (athermal system). Generalizing this model, by including self-attraction in the chain, the globule phase as well as a coil-globule transition arise. When the polymer is placed on a lattice, the standard interacting SAW (ISAW) model consists in assigning an energy $\epsilon$ (yielding an attractive force) between monomers on \textit{nearest-neighbor} (NN) sites nonconsecutive in the walk \cite{Carlo,Flory2,DeGennes2}. Coil and globule phases, separated by a tricritical ($\theta$)-point, are indeed observed in this model. In two-dimensions, the exponents $\nu_{coil}=3/4$, $\nu_{\theta}=4/7$ and $\nu_{globule}=1/2$ are exactly known \cite{Flory1,DupSaleur,Carlo}. The more general case of semiflexible polymers has been modeled by introducing a bending energy $\epsilon_b$ in the ISAW model (see for instance Refs. \cite{jurgenSF1,Lise,pretti,Zhou2}). In this semiflexible-ISAW (sISAW) model, a stable crystalline (solid-like) phase also exists in the system (for low $T$ and large $\epsilon_b$), in addition to the coil and globule ones.

Lee \textit{et al.} \cite{Lee1,Lee2} have proposed another interesting generalization of the ISAW model by associating different energies $\epsilon_1$ and $\epsilon_2$ to monomers on NN and \textit{next}-NN (NNN) sites, respectively. From exact enumeration of walks with up to 38 monomers on the square lattice, a line of $\theta$-points (a $\theta$-line) separating the coil and globule phases was found. Similar results were also observed in recent Monte Carlo simulations of this model on the square and cubic lattices \cite{Nathann}. Interestingly, this last study showed that the $\theta$-line exists even for competing interactions between monomers ($\epsilon_1 < 0$ and $\epsilon_2>0$ or $\epsilon_1 > 0$ and $\epsilon_2 < 0$). Actually, a $\theta$-line and the absence of other phases (beyond the coil and globule ones) is quite expected when both forces are attractive ($\epsilon_1>0$ and $\epsilon_2>0$), but for competing interactions this is not necessarily the case, due to the frustration arising from such competition, which might change the critical properties of the system. As a classical example of this, one may cite the Ising model on the square lattice with competing NN and NNN interactions, where different ordered phases, transitions and universality classes are observed (for a recent survey see \cite{Sandvik}). In polymers, competing (on-site) interactions in the multiple monomer per site (MMS) model by Krawczyk \textit{et al.} \cite{Krawczyk} is known to change the coil-globule transition in a certain region of its phase diagram, but the order of transition still remain unclear \cite{Tiago}.

Another interesting feature of the ISAW model when the force between NNN monomers is repulsive is its semiflexibility, because $\epsilon_{2}<0$ acts as a bending energy, though it also repels NNN monomers that are not part of a bend. Anyway, for large enough $\epsilon_1>0$ and $\epsilon_2<0$, a crystalline phase could be expected in this model. However, at least in the range of energies analyzed in Ref. \cite{Nathann}, this crystalline phase was not observed.

In order to analyze in more detail whether competing forces between monomers can change or not the coil-globule transition, as well as whether it yields or not an ordered (crystalline) phase in the ISAW with NNN interactions, here, we solve this model on a Husimi lattice built with squares. Different definitions of next nearest-neighbors (and interactions between them) on this lattice are analyzed, but in all cases the same qualitative results are obtained: no crystalline phase is found and the $\theta$-line extends over the whole phase diagram, for $\epsilon_1$ in the range $[-\infty, \infty)$, which includes the regions of competing interactions. Only in the extremal case of an infinite repulsive force between NNN monomers there is a breakdown of the coil-globule transition, which is quite expected since in this case the chains are straight.

The rest of this work is organized as follows. In Sec. \ref{defmod} the model is defined on a Husimi lattice built with squares and solved in terms of recursion relations. The thermodynamic properties of the model are presented in Sec. \ref{tpmA}. In Sec. \ref{conclusions} our final discussions and conclusions are summarized. The calculations of the free energy and of the $\theta$-lines are demonstrated in appendices \ref{ApFreeEnergy} and \ref{ApTricrical}, respectively.

\section{Definition of the model and its solution in terms of recursion relations}
\label{defmod}

\begin{figure}[!t]
\begin{center}
\includegraphics[width=13cm]{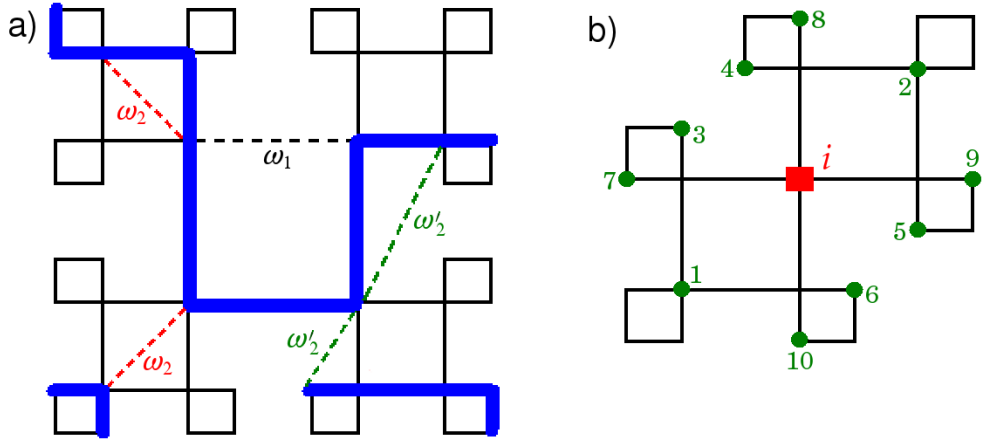}
\caption{(Color online) a) Example of contribution to the partition function of the model on a Husimi tree with three generations. The polymer chains are represented by full thick (blue) lines and the dashed lines gives examples of each type of monomer interaction. The weight of this configuration is $z^{18} \omega_{1}^{2} \omega_{2}^{8} \omega_{2}'^{4}$. b) Definition of the second neighbor (circles) - by the chemical distance - of the site $i$ (square).}
\label{FigRede}
\end{center}
\end{figure}

We investigate interacting self- and mutually avoiding walks on a Husimi lattice - the core of a Cayley tree \cite{b82} - built with squares (see Fig.\ref{FigRede}). The endpoints of the walks are placed on the surface of the tree. In our grand-canonical solution, the thermodynamic variables of interest are the monomer fugacity $z$ and the Boltzmann weights $\omega_{1}=\exp(\epsilon_1/k_B T)$ and $\omega_{2}=\exp(\epsilon_2/k_B T)$ associated to each pair of nonconsecutive monomers on nearest-neighbor (NN) sites and pairs of monomers on next-NN (NNN) sites on the lattice, respectively. Hereafter, we will refer to them as NN and NNN monomers. Then, the grand-canonical partition function of the model is given by
\begin{equation}
Y=\sum z^{M} \omega_1^{M_{NN}}\omega_2^{M_{NNN}},
\end{equation}
where the sum runs over all configurations of the walks on the tree, and $M$, $M_{NN}$ and $M_{NNN}$ are respectively the total number of monomers and the number of NN and NNN monomers. At this point, one notices that on the Husimi lattice there exists an ambiguity in definition of NNN sites:

\textit{a}) the neighborhood of a given site, let us say $i$, can be defined according to the chemical distance (associated to the number of steps along the lattice edges to reach a site $j$ starting at $i$). So, at first, any site would have ten second neighbors, as shown in Fig. \ref{FigRede}b. However, four of these sites (7-10 in Fig. \ref{FigRede}b) would correspond to third neighbors on the square lattice. Since our aim is to compare the Husimi solution with results for the ordinary lattice, it is more reasonable to define only six NNN sites (the 1-6 ones in Fig. \ref{FigRede}b).

\textit{b}) Another option is to state that second neighbors are the sites in opposite vertices of an elementary square (sites 1-$i$ and 2-$i$, in Fig. \ref{FigRede}b). Then, each site will have two NNN ones.

Since definition \textit{a} (\textit{b}) - hereafter called approach A (C) - overestimates (underestimates) the number of second neighbors on square lattice - which is four -, both approaches will be analyzed in the following. It is easy to see in Fig. \ref{FigRede}b that the overestimate in case A comes from the ``out-square'' sites 3-6, because in the ordinary lattice 3 and 4 (and also 5 and 6) would be a single site. So, this ``excess'' of second neighbors can be compensated by assigning only half of NNN energy ($\epsilon_2/2$) for ``out-square'' NNN monomers. In this way, the contribution to the partition function of the possible six NNN monomers will be effectively the same of four NNN ones (i.e., $\omega_2^4$). This case will be referred to as approach B. In order to consider all these cases in a general way, one may assign a weight $\omega_2$ for ``in-square'' NNN monomers and a weight $\omega_2'$ for the ``out-square'' NNN monomers (see Fig. \ref{FigRede}a). Thence, the approaches A, B and C are recovered by making $\omega_2'=\omega_2$, $\omega_2'=\sqrt{\omega_2}$ and $\omega_2'=1$, respectively.

\begin{figure}[t]
\begin{center}
\includegraphics[width=10.cm]{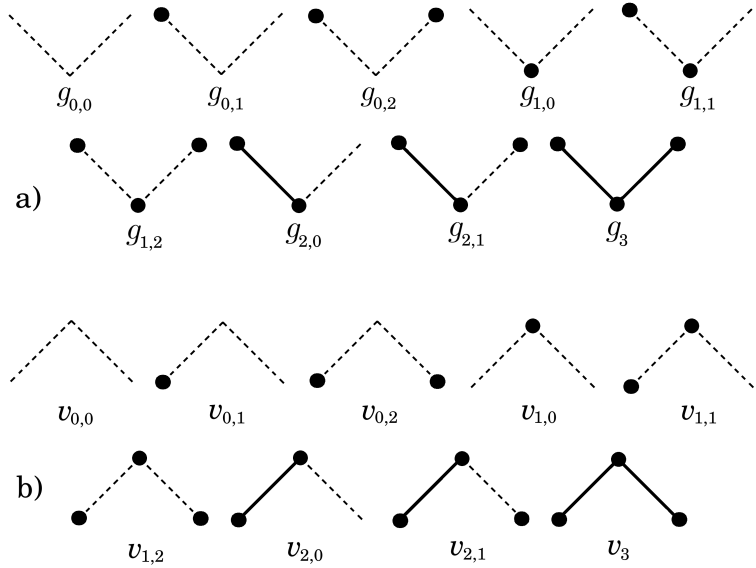}
\caption{Definition of a) the root sites for each partial partition function and b) the types of possible vertices. Circles indicate the presence of monomers in the vertex and their bonds are represented by full lines.}
\label{FigRoots}
\end{center}
\end{figure}

A subtree with generation $M+1$ can be obtained by attaching 3 subtrees (each one with $M$ generations) in 3 vertices of an elementary square. The remaining vertex is usually called the ``root site'' and, associated to it, a partial partition function (ppf) $g_i$ is defined. Nine root sites and, thus, nine ppf's are required to correctly account for the NNN interactions (see Fig. \ref{FigRoots}a). The ppf $g_i$ in generation $M+1$ is determined counting all possible configurations produced by attaching 3 subtrees (each one with $M$ generations) in a square with root site of type $i$. Instead of do this directly, it is convenient to determine first the contribution coming from each type of vertex of the elementary square (depicted in Fig. \ref{FigRoots}b), being

\begin{subequations}
\label{EqVert}
\begin{equation}
v_{0,0} = g_{0,0} + 2 g_{0,1} + g_{0,2},
\end{equation}
\begin{equation}
v_{0,1} = g_{0,0} + (1 + \omega_2') g_{0,1} + \omega_2' g_{0,2},
\end{equation}
\begin{equation}
v_{0,2} = g_{0,0} + 2 \omega_2' g_{0,1} + \omega_2'^2 g_{0,2},
\end{equation}
\begin{equation}
v_{1,0} = z g_{3},
\end{equation}
\begin{equation}
v_{1,1} = z \omega_2' g_{3},
\end{equation}
\begin{equation}
v_{1,2} = z \omega_2'^2 g_{3},
\end{equation}
\begin{equation}
v_{2,0} = z [(1+\omega_2') g_{2,0} + 2 \omega_2' g_{2,1}],
\end{equation}
\begin{equation}
v_{2,1} = 2 z (\omega_2' g_{2,0} + \omega_2'^2 g_{2,1}),
\end{equation}
\begin{equation}
v_{3} = z (g_{1,0} + 2 \omega_2' g_{1,1} + \omega_2'^2 g_{1,2}).
\end{equation}
\end{subequations}

In terms of these expressions, it is quite simple to determine the recursion relations for the ppf's of the model, given by

\begin{subequations}
\label{EqPPFs}

\begin{equation}
g_{0,0}'=v_{0,0}^3 + v_{0,1}^2 v_{1,0},
\end{equation}
\begin{equation}
g_{0,1}'=v_{0,0} v_{0,1} v_{1,0} + \omega_1 v_{0,1} v_{1,1}^2 + v_{0,1} v_{2,0}^2,
\end{equation}
\begin{equation}
g_{0,2}'=\omega_2 [v_{0,2} v_{1,0}^2 + \omega_1^2 v_{1,1}^2 v_{1,2} + 2 \omega_1 v_{1,1} v_{2,0} v_{2,1} + v_{2,0}^2 v_{3}],
\end{equation}
\begin{equation}
g_{1,0}'=v_{0,0} v_{0,1}^2 + \omega_{2} v_{0,2}^2 v_{1,0},
\end{equation}
\begin{equation}
g_{1,1}'= \omega_1 [v_{0,1}^2 v_{1,1} + \omega_1 \omega_2 v_{0,2} v_{1,1} v_{1,2} + \omega_2 v_{0,2} v_{2,0} v_{2,1}],
\end{equation}
\begin{equation}
g_{1,2}'= \omega_1^2 \omega_2 [v_{0,2} v_{1,1}^2 + \omega_1^2 \omega_2 v_{1,2}^3 + 2 \omega_1 \omega_2 v_{1,2} v_{2,1}^2 + \omega_2 v_{2,1}^2 v_{3}],
\end{equation}
\begin{equation}
g_{2,0}'= v_{0,1}^2 v_{2,0} + \omega_1 \omega_2 v_{0,2} v_{1,1} v_{2,1} + \omega_2 v_{0,2} v_{2,0} v_{3},
\end{equation}
\begin{equation}
g_{2,1}'= \omega_1 \omega_2 [v_{0,2} v_{1,1} v_{2,0} + \omega_1^2 \omega_2 v_{1,2}^2 v_{2,1} + \omega_1 \omega_2 (v_{2,1}^3 + v_{1,2} v_{2,1} v_{3}) + \omega_2 v_{2,1} v_{3}^2],
\end{equation}
\begin{equation}
g_{3}'= \omega_2 [v_{0,2} v_{2,0}^2 + \omega_1^2 \omega_2 v_{1,2} v_{2,1}^2 + 2 \omega_1 \omega_2 v_{2,1}^2 v_{3}],
\end{equation}
 
\end{subequations}
where $g_i$ and $g_i'$ are in generations $M$ and $M+1$, respectively.

In a similar way, the partition function of the model on the Cayley tree can be find by attaching four subtrees in a central square, which yields
\begin{eqnarray} 
\label{EqFuncPart1}
Y&=& v_{0,0}^4 + 4 v_{0,0} v_{0,1}^2 v_{1,0} + 4 \omega_1 v_{0,1}^2 v_{1,1}^2 + 2 \omega_2 v_{0,2}^2 v_{1,0}^2 + 4 \omega_1^2 \omega_2 v_{0,2} v_{1,1}^2 v_{1,2} + \omega_1^4 \omega_2^2 v_{1,2}^4 + 4 v_{0,1}^2 v_{2,0}^2 \\ \nonumber
&+& 8 \omega_1 \omega_2 v_{0,2} v_{1,1} v_{2,0} v_{2,1}  + 4 \omega_2 v_{0,2} v_{2,0}^2 v_3 + 4 \omega_1^3 \omega_2^2 v_{1,2}^2 v_{2,1}^2 + 2 \omega_1^2 \omega_2^2  (v_{2,1}^4 + 2 v_{1,2} v_{2,1}^2 v_3) + 4 \omega_1 \omega_2^2 v_{2,1}^2 v_3^2.
\end{eqnarray}
Then, the densities of monomers ($\rho$), of NN ($\rho_{NN}$) and NNN ($\rho_{NNN}$) monomers are
\begin{equation}
 \rho = \frac{z}{4 Y} \left( \frac{\partial Y}{\partial z}\right), \quad \quad \rho_{NN} = \frac{\omega_1}{4 Y} \left(\frac{\partial Y}{\partial \omega_1}\right) \quad \quad \text{and} \quad \quad \rho_{NNN} = \frac{\omega_2}{S Y} \left(\frac{\partial Y}{\partial \omega_2} \right),
\label{EqDensities}
\end{equation}
where $S=10$, $6$ and $2$ are used in approaches A, B and C, respectively, to make the maximum value of $\rho_{NNN}$ equal 1 in all cases.

In the thermodynamic limit, when the number of generations of the tree and, consequently, the length of the polymers tend to infinity, the ppf's diverge, so, we will work with ratios of them, defined as $R_{1} = g_{0,1}/g_{0,0}$, $R_{2} = g_{0,2}/g_{0,0}$, $R_{3} = g_{1,0}/g_{0,0}$, $R_{4} = g_{1,1}/g_{0,0}$, $R_{5} = g_{1,2}/g_{0,0}$, $R_{6} = g_{2,0}/g_{0,0}$, $R_{7} = g_{2,1}/g_{0,0}$ and $R_{8} = g_{3}/g_{0,0}$. This leads to the recursion relations (RR's):
\begin{subequations}
\label{EqRRsA}
\begin{equation}
R_1' = (z A B R_8 + z^2 \omega_1 \omega_2^{2\alpha} B R_8^2 + z^2 B D^2)/R_0,
\end{equation}
\begin{equation}
R'_2 = \omega_2 (z^2 C R_8^2 + z^3 \omega_1^2 \omega_2^{4\alpha} R_8^3 + 4 z^3 \omega_1 \omega_2^{\alpha} D E R_8 + z^3 D^2 F)/R_0,
\end{equation}
\begin{eqnarray}
R'_3 = (A B^2 + z \omega_2 C^2 R_8)/R_0,
\nonumber
\end{eqnarray}
\begin{equation}
R'_4 = \omega_1 (z \omega_2^{\alpha} B^2 R_8 + z^2 \omega_1 \omega_2^{3\alpha+1} C R_8^2 + 2 z^2 \omega_2 C D E)/R_0,
\end{equation}
\begin{equation}
R'_5 = \omega_1^2 \omega_2 (z^2 \omega_2^{2\alpha} C R_8^2 + z^3 \omega_1^2 \omega_2^{6\alpha+1} R_8^3 + 8 z^3 \omega_1 \omega_2^{2\alpha+1} E^2 R_8 + 4 z^3 \omega_2 E^2 F)/R_0,
\end{equation}
\begin{equation}
R'_6 = (z B^2 D + 2 z^2 \omega_1 \omega_2^{\alpha+1} C E R_8 + z^2 \omega_2 C D F)/R_0,
\end{equation}
\begin{equation}
R'_7 = \omega_1 \omega_2 [z^2 \omega_2^{\alpha} C D R_8 + 2 z^3 \omega_1^2 \omega_2^{4\alpha+1} E R_8^2 + \omega_1 \omega_2 (8 z^3 E^3 + 2 z^3 \omega_2^{2\alpha} E F R_8 ) + 2 z^3 \omega_2 E F^2]/R_0,
\end{equation}
\begin{equation}
R'_8 = \omega_2 (z^2 C D^2 + 4 z^3 \omega_1^2 \omega_2^{2\alpha+1} E^2 R_8 + 8 z^3 \omega_1 \omega_2 E^2 F)/R_0,
\end{equation}
with
\begin{equation}
R_0=(1+2 R_1+R_2)^3 + z [1+(1+\omega_2^{\alpha}) R_1 + \omega_2^{\alpha} R_2]^2  R_8,
\end{equation}
\begin{equation}
A = 1 + 2 R_1 + R_2,
\end{equation}
\begin{equation}
B = 1+(1+\omega_2^{\alpha}) R_1 + \omega_2^{\alpha} R_2,
\end{equation}
\begin{equation}
C = 1 + 2 \omega_2^{\alpha} R_1 + \omega_2^{2\alpha} R_2,
\end{equation}
\begin{equation}
D = (1 + \omega_2^{\alpha}) R_6 + 2 \omega_2^{\alpha} R_7,
\end{equation}
\begin{equation}
E = \omega_2^{\alpha} R_6 + \omega_2^{2\alpha} R_7,
\end{equation}
\begin{equation}
F = R_3 + 2 \omega_2^{\alpha} R_4 + \omega_2^{2\alpha} R_5,
\end{equation}
\end{subequations}
where $\alpha=1$, $1/2$ and $0$ in cases A, B and C, respectively. In the case C ($\omega_2'=1$), it is easy to see that $v_{0,0} = v_{0,1} = v_{0,2} = (g_{0,0} + 2 g_{0,1} + g_{0,2})$, $v_{1,0} = v_{1,1} = v_{1,2} = z g_{3}$, $v_{2,0} = v_{2,1} = 2 z (g_{2,0} + g_{2,1})$ and $v_{3} = z (g_{1,0} + 2 g_{1,1} + g_{1,2})$ and, thus, only the combinations $g_0 \equiv (g_{0,0} + 2 g_{0,1} + g_{0,2})$, $g_1 \equiv (g_{1,0} + 2 g_{1,1} + g_{1,2})$ and $g_2 \equiv (g_{2,0} + g_{2,1})$ will appear in the recursion relations. Thus, one may work with simplified ratios of ppf's, defined as $R_1=g_1/g_0$, $R_2=g_2/g_0$ and $R_3=g_3/g_0$, yielding
\begin{subequations}
\label{EqRRsB}
\begin{equation}
R_{1}'=[1 + z (2 \omega_{1}+\omega_{2}) R_3 + 3 z^2 \omega_{1}^2 \omega_{2} R_3^2 + z^3 \omega_{1}^4 \omega_{2}^2 R_3^3 + 8 z^2 \omega_{1} \omega_{2} R_2^2 + 8 z^3 \omega_{1}^3 \omega_{2}^2 R_2^2 R_3 + 4 z^3 \omega_{1}^2 \omega_{2}^2 R_{1} R_2^2 ]/R_0,
\end{equation}
\begin{equation}
R_{2}'= 2 z R_2 (1 + 2 z \omega_{1} \omega_{2} R_3 + z^2 \omega_{1}^3 \omega_{2}^2 R_3^2 + 4 z^2 \omega_{1}^2 \omega_{2}^2 R_2^2 + z \omega_{2} R_{1} + z^2 \omega_{1}^2 \omega_{2}^2 R_{1} R_3 + z^2 \omega_{1} \omega_{2}^2 R_{1}^{2} )/R_0,
\end{equation}
\begin{equation}
 R_{3}'= 4 z^2 R_2^2 \omega_{2} (1 + z \omega_{1}^2 \omega_{2} R_3 + 2 z \omega_{1} \omega_{2} R_{1} )/R_0,
\end{equation}
with
\begin{equation}
R_0 = 1 + 3 z R_3 + z^2 (2 \omega_{1}+\omega_{2}) R_3^2 + z^3 \omega_{1}^2 \omega_{2} R_3^3 + 8 z^2 R_2^2 + 8 z^3 \omega_{1} \omega_{2} R_2^2 R_3 + 4 z^3 \omega_{2} R_{1} R_2^2 
\end{equation}
\end{subequations}

It is worth noting that the partition functions (Eq. \ref{EqFuncPart1}) for approach $X$, with $X=$A, B or C, can be written as $Y_{X}=g_{0,0}^4 y_{X}$, where $y_{X}$ is finite (since is depends only on the ratios $R_i$, beyond $z$, $\omega_1$ and $\omega_2$), while $Y_{X}$ diverges as $g_{0,0}^4$, in the thermodynamic limit. Notwithstanding, the densities (Eq. \ref{EqDensities}) remain finite because they are in fact functions of $y_{X}$, instead of $Y_{X}$.

\section{Thermodynamic properties of model} 
\label{tpmA}

The thermodynamic phases of the model on the Husimi lattice are given by the real and positive fixed points of the recursion relations (RR's - Eqs. \ref{EqRRsA}). Similarly to the classical ISAW model ($\omega_2=1$), the grand-canonical phase diagram for general $\omega_2$ presents only two phases: \textit{i)} a non-polymerized (NP) phase, where $R_3=1$ and $R_i=0$ otherwise; and \textit{ii)} a polymerized (P) phase, with $R_i \neq 0$ for $i=1, \ldots, 8$. In the former, the density of monomers vanishes ($\rho=0$) and, consequently, $\rho_{NN}=\rho_{NNN}=0$. On the other hand, in P phase these densities are, in general, non-null and depend on the parameters $z$, $\omega_{1}$ and $\omega_{2}$. Obviously, working with the reduced set of RR's (Eqs. \ref{EqRRsB}) in approach C, one finds a similar behavior, with $R_1=1$, $R_2=R_3=0$ in NP phase, and $R_i \neq 0$ in P one.

Each phase is stable in the region of the parameter space ($z,\omega_1,\omega_2$) where the largest eigenvalue $\lambda$ of its Jacobian matrix $\left( J_{i,j}=\frac{\partial R'_i}{\partial R_j}\right) $ is smaller than one. The condition $\lambda=1$ gives the thermodynamic stability limit (the spinodal) of the respective phase, which is easy to calculate in NP phase, being
\begin{equation}
\omega_1 = \frac{1}{2 z^3 \omega_2^{2(\alpha+1)}} \left( \frac{-1 + z + z \omega_2^{\alpha} + z^2 \omega_2 + z^2 \omega_2^{\alpha+1}}{-1 - z + z \omega_2^{\alpha} - z^2 \omega_2 + z^2 \omega_2^{\alpha+1}} \right),
\label{EqNPlimitA}
\end{equation}
recalling that $\alpha=1$, $1/2$ and $0$ in approaches A, B and C, respectively. For the polymerized phase, the stability limit is determined numerically. In a certain region of the phase diagram, the NP and P spinodals are coincident, forming a critical surface.

There exists also a coexistence region in the phase diagram - where the spinodals do not match and both phases are stable - with a coexistence surface separating the NP and P phases there. A simple way to determine this surface is through the free energy of the model, which can be calculated using Gujrati's prescription \cite{Gujrati}. The derivation of this free energy for the Husimi lattice built with squares is presented in appendix \ref{ApFreeEnergy}, leading to
\begin{equation}
 \phi_{b} = -\dfrac{1}{2} \left( 2 \ln R_{0,X} - \ln y_{X} \right)
\end{equation}
with $X=$A, B or C, and $R_{0,X}$ and $y_X$ defined as above and calculated at the fixed point. In NP phase, $R_{0,X}=y_X=1$, so that $\phi_{b}^{NP}=0$ and, then, the coexistence surface - where the free energies of both phases are equal - is given by $\phi_{b}^{P}=\phi_{b}^{NP}=0$.

These critical and coexistence surfaces meet (tangentially) at a tricritical (TC) line, which was calculated exactly by locating the points along the spinodal of NP phase where the solution of the RR's is triply degenerate. This is demonstrated in detail in appendix \ref{ApTricrical}.

\begin{figure}
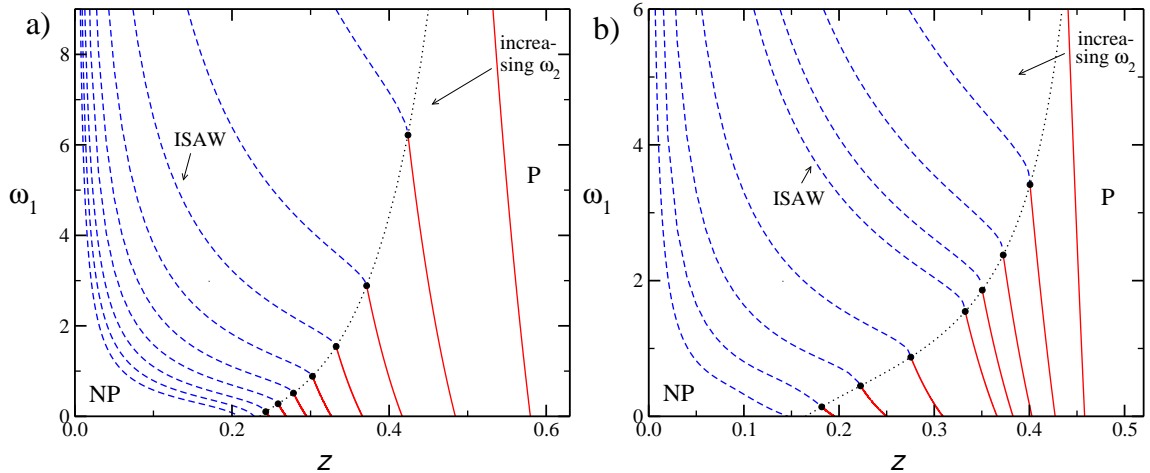

\begin{center}
\includegraphics[width=7.5cm]{DiagramasA.eps}
\includegraphics[width=7.5cm]{DiagramasC.eps}
\caption{(Color online) Phase diagrams in the variables $\omega_1$ versus $z$ for approaches a) A and b) C. In a) diagrams for $\omega_2$ varying by $0.2$ in the range $[0.4,2.2]$ $\omega_2=0.4,0.6,0.8,1.0,1.6,2.2,2.8,3.4$ and $4.0$ are shown, while in b) they are for $\omega_2=0.2,0.4,0.6,0.8,1.0,2.0,4.0,8.0$ and $16.0$. The full (red) and dashed (blue) lines are the critical and coexistence lines, respectively. The black dots indicate the $\theta$ points and the $\theta$ line is given by the dotted (black) line.}
\label{FigDiagsAC}
\end{center}
\end{figure}

Let us first discuss the thermodynamic behavior for approach A. Figure \ref{FigDiagsAC}a shows phase diagrams for several values of (fixed) $\omega_2$ in the range $[0.4,2.2]$. For any $\omega_2 < 1.9217692$, the same properties of the classical ISAW model ($\omega_2=1$) are found: there is a continuous NP-P transition at small $\omega_1$ that becomes discontinuous at a tricritical ($\theta$) point. The location of the $\theta$ point, however, strongly depends on $\omega_2$, forming a continuous $\theta$-line. 

In general, a decrease in the coordinates ($z^{\theta},\omega_1^{\theta}$) of the tricritical point is observed as $\omega_2$ increases. In fact, for attractive forces between NN and NNN monomers, this is quite expected, since $\beta\epsilon_{2}>0$ will facilitate the collapse and, thus, $\beta\epsilon_{1}$ becomes smaller. When $\omega_2=1.1553956$, the NN energy is null, i.e., $\omega_1^{\theta}=1$ (and $z^{\theta}=0.3085453$), so that the collapse transition happens due solely to the NNN interaction. For larger $\omega_2$, the $\theta$-line still exists, but for $\omega_1^{\theta}<1$, meaning that NN monomers repel each other. The value of $\omega_1^{\theta}$ decreases, for increasing $\omega_2$, until reaches the zero at $\omega_2 = 1.9217692$. Therefore, even for an infinite repulsive force between NN monomers a coil-globule transition exists for a finite (attractive) interaction between NNN ones. This will be discussed in more detail in the following. For $\omega_2 > 1.9217692$, only a NP-P coexistence surface is found.

The $\theta$-line extends also to the region of repulsive NNN interactions ($\omega_2<1$), where, again, increasing $z^{\theta}$ and $\omega_1^{\theta}$ are observed as $\omega_2$ decreases. When $\omega_2$ approaches the zero, one finds $\omega_1^{\theta} \rightarrow \infty$ and $z^{\theta} \rightarrow 1$. This is quite reasonable, since $\omega_2 \ll 1$ will prevent the formation of bends in the walks and, consequently, of a globular phase. Notwithstanding, if $\omega_1 \gg 1$ the attractive NN force can overcome the NNN repulsion yielding this phase.

\begin{figure}
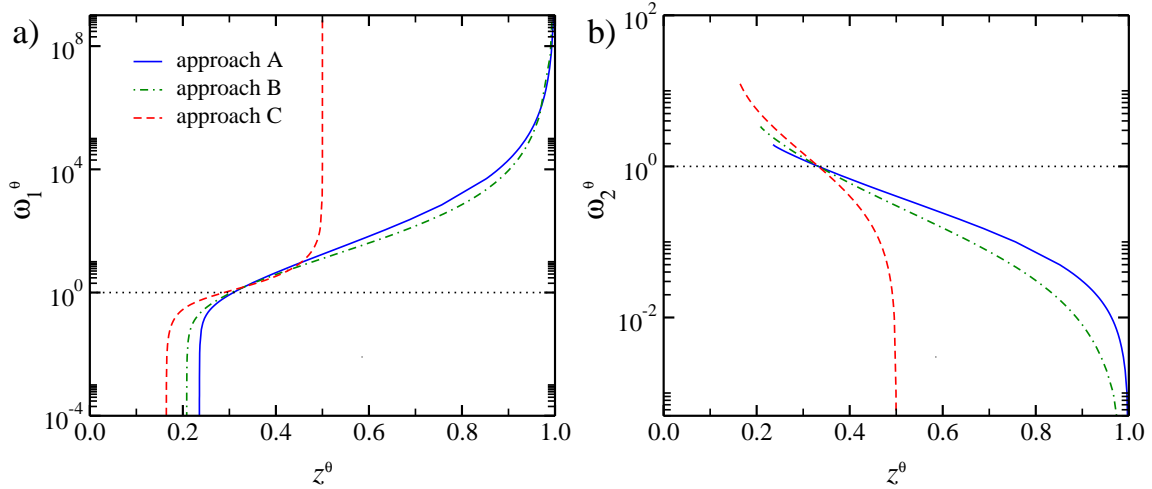

\begin{center}
\includegraphics[width=7.50cm]{DiagThetaLinesw1.eps}
\includegraphics[width=7.50cm]{DiagThetaLinesw2.eps}
\caption{(Color online) Values of a) $\omega_{1}^{\theta}$ and b) $\omega_{2}^{\theta}$ against $z^\theta$, for all approaches. The horizontal (dotted, black) line separates the regions where interactions are attractive and repulsive.}
\label{FigDiagThetaAB}
\end{center}
\end{figure}

The phase diagram for approach C is presented in Fig. \ref{FigDiagsAC}b, where the same qualitative behavior of A is observed, with $\theta$-lines extending also over the whole phase diagrams since $\omega_1=0$ until $\omega_1 \rightarrow \infty$. An analogous phase diagram is also found for the intermediate approach B (not shown). These similarities are more evident in the comparison of the $\theta$-lines, for all approaches, which is presented in Fig. \ref{FigDiagThetaAB}. In case A (B), the $\theta$-line starts at $z^{\theta} = 0.235592$ ($0.208304$) - where $\omega_{1}^{\theta} = 0$ and $\omega_{2}^{\theta} = 1.921769$ ($3.345581$) - and ends at $z^{\theta} \rightarrow 1$ - where $\omega_{1}^{\theta} \rightarrow \infty$ and $\omega_{2}^{\theta} \rightarrow 0$. On the other hand, in case C a quite different $z$-range is found for the $\theta$-line, which starts at $z^{\theta} = 0.16447819$ - where $\omega_{1}^{\theta} = 0$ and $\omega_{2}^{\theta} = 12.4023526$ - and ends at $z^{\theta} \rightarrow 1/2$ - where $\omega_{1}^{\theta} \rightarrow \infty$ and $\omega_{2}^{\theta} \rightarrow 0$. Notice that $\omega_2^{\theta} (\text{C}) > \omega_2^{\theta} (\text{B}) > \omega_2^{\theta} (\text{A})$ for attractive NNN interactions and $\omega_2^{\theta} (\text{C}) < \omega_2^{\theta} (\text{B}) < \omega_2^{\theta} (\text{A})$ for repulsive ones is quite expected, since the effective number of possible NNN monomers decreases from A to C.

We notice that in approach C a repulsive (attractive) NNN interaction introduces energetic penalties (advantages) whenever the polymer bending within an elementary square, but do not when it bends in the opposite direction. This unrealistic feature of the Husimi lattice in case C certainly explains why $z^{\theta} \rightarrow 1$ (in cases A and B) and $z^{\theta} \rightarrow 1/2$ (in C), when $\omega_2^{\theta} \rightarrow 0$ (with $\omega_1^{\theta} \rightarrow \infty$).

It is important to remark that, when NNN monomers repel each other, the polymer is semiflexible and, thus, an anisotropic/crystalline phase could be expected in the phase diagrams, for large enough $\omega_1$. However, we have exhaustively looked for any new stable phase in this region and did not find any. One recalls that the crystalline phase is dense ($\rho \approx 1$) - it is a quasi-Hamiltonian walk - featured by straight parallel chains, maximizing the number of NN monomers and minimizing the bending. Therefore, to correctly analyze this phase - with chains aligned in one direction of the lattice -, at first, more general RR's are required, defining the root sites (and ppf's) to account for the directional anisotropy (as done for the sISAW in \cite{pretti}, for example). In the homogeneous solution, analyzed here, both directions are treated as equivalent (in the same ppf) and, thus, the symmetry-breaking of the phase cannot arise. In any case, however, a dense phase should appear as a diverging fixed point of the RR's, because they were defined as $R_i = g_i/g_{0,0}$ and configurations of type $g_{0,0}$ (see Fig. \ref{FigRoots}) do not exist in a fully occupied lattice. Thus, although we are analyzing only the homogeneous case, the absence of a divergence in the RR's strongly suggests that there is no crystalline phase in the model on the Husimi lattice. In fact, in contrast to the bending energy in the sISAW model, in our case the (repulsive) force acts on NNN monomers regardless they are part of a bending and, thus, in such crystalline phase the (repulsive) NNN interaction would be also maximized, together with the (attractive) NN one. This frustration in the system is certainly responsible for the absence of the order.

\subsection{Infinite repulsion between NN monomers ($\omega_{1}=0$)}

\begin{figure}[t]
\begin{center}
\includegraphics[width=10.0cm]{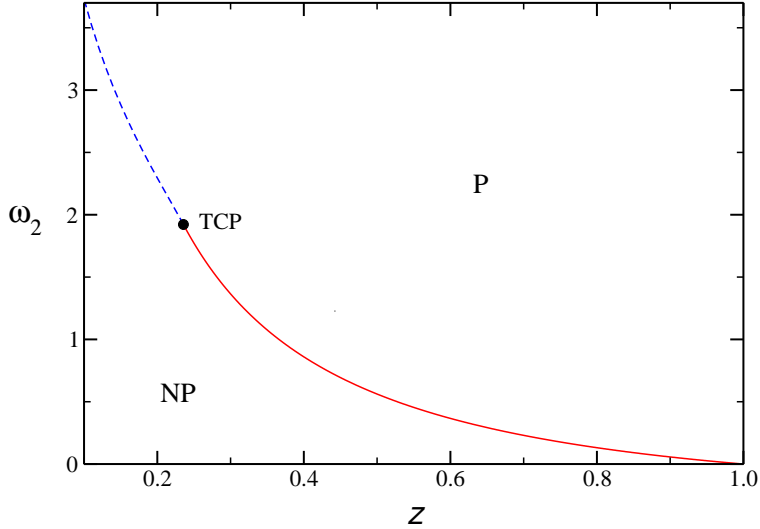}
\caption{(Color online) Phase diagram for approach A and $\omega_1=0$. The full (red) and dashed (blue) lines are the critical and coexistence lines, respectively. The black dot indicate the tricritical point.}
\label{FigDiag_w1_000}
\end{center}
\end{figure}

Now, we consider the case where NN monomers are forbidden ($\omega_1=0$). As pointed above, phase diagrams similar to the ones for finite NN interaction are found also in this limit. Indeed, from equation \ref{EqNPlimitA}, the NP spinodal can be written as
\begin{equation}
z=\frac{-1-\omega_2^{\alpha}+\sqrt{1+2 \omega_2^{\alpha}+\omega_2^{2\alpha}+4 \omega_2+4 \omega_2^{\alpha+1}}}{2(\omega_2+\omega_2^{\alpha+1})}.
\label{EqLimNPw10}
\end{equation}
For $\omega_2 < \omega_{2}^{\theta}$, this expression defines also the critical line. Once more, $\alpha=1$ (in case A), $\alpha=1/2$ (in B) and $\alpha=0$ (in C). From the analysis in appendix \ref{ApTricrical}, the values of $\omega_{2}^{\theta}$ are given by the real positive root of the polynomial
\begin{eqnarray}
&7&+29 \omega_2+26 \omega_2^{\alpha}-2 \omega_2^2+74 \omega_2^{\alpha+1}+33 \omega_2^{2\alpha}-6 \omega_2^{2(\alpha+1)}+12 \omega_2^{3\alpha}-7 \omega_2^{4\alpha}-6 \omega_2^{5\alpha}-\omega_2^{6\alpha} \\ \nonumber
&+&46 \omega_2^{2\alpha+1}-2 \omega_2^{3\alpha+2}-2 \omega_2^{5\alpha+1}-16 \omega_2^{3\alpha+1}-19 \omega_2^{4\alpha+1}-6 \omega_2^{\alpha+2}+b(\omega_2^{5\alpha}+\omega_2^{4\alpha}+\omega_2 +1 \\  \nonumber
&-&2 \omega_2^{3\alpha}+3 \omega_2^{\alpha+1}-2 \omega_2^{2\alpha}+\omega_2^{\alpha}+\omega_2^{3\alpha+1}+3 \omega_2^{2\alpha+1})=0,
\end{eqnarray}
with $b=\sqrt{1+2 \omega_2^{\alpha}+\omega_2^{2\alpha}+4 \omega_2+4 \omega_2^{\alpha+1}}$.
Actually, in each approach, there are two of such roots, but one of them leads to inconsistent values of $z$, being $\omega_{2}^{\theta} = 1.9217693$ (A), $\omega_{2}^{\theta} = 3.3455816$ (B) and $\omega_{2}^{\theta} = 12.4023526$ (C) the physical ones. Inserting these values in Eq. \ref{EqLimNPw10} one readily finds $z^{\theta} = 0.2355927$ (A), $z^{\theta} = 0.2083038$ (B) and $z^{\theta} = 0.1644782$ (C). In the region $z < z^{\theta}$ (where $\omega_2 > \omega_{2}^{\theta}$) both phases coexist. One example of this phase behavior is presented in Fig. \ref{FigDiag_w1_000} for case A and analogous ones are obtained in other approaches (not shown).

Although a coil-globule transition is present in this case, the globule phase is different from the one for $\omega_1 >0$, since $\omega_1=0$ forbids the presence of NN monomers in the system (i.e., $\rho_{NN}=0$). For instance, from the expressions for the densities of monomers $\rho$ and NNN monomers $\rho_{NNN}$ (not shown explicitly here), it is possible to demonstrate that $\rho \rightarrow 1$ and $\rho_{NNN} \rightarrow 1$, in the limit $\omega_2 \rightarrow \infty$, for any approach when $\omega_1 > 0$. However, for $\omega_1=0$, a diverging $\omega_2$ leads to $\rho \rightarrow 3/4$ and $\rho_{NNN} \rightarrow 7/10$ (in case A), $\rho \rightarrow 3/4$ and $\rho_{NNN} \rightarrow 2/3$ (in B), and $\rho \rightarrow 2/3$ and $\rho_{NNN} \rightarrow 1/2$ (in C), which corresponds to ``soft'' polymerized phases (and respective ``soft globules''), since the maximal occupation of the lattice is smaller than one. In a canonical situation (of polymers with fixed size), this ``soft'' phase (for $\omega_1 = 0$) shall occupy a volume larger than the ``regular'' globule phase (for $\omega_1 > 0$).

We claim that this ``soft'' globule phase is not a feature of the Husimi lattice, but it might exist also in the square (and other regular) lattices. In fact, although NN monomers are forbidden, the attractive force between NNN monomers that are not part of a bending can acts to collapse the chains, in a similar way as the NN one in the ordinary ISAW model. Moreover, the NNN interaction enhances the formation of bends and, consequently, the formation of globules. Anyhow, more studies are necessary to confirm this.

\subsection{Infinite repulsion between NNN monomers ($\omega_{2}=0$)}

Now, we turn our attention for the case of forbidden NNN monomers. Making $\omega_2=0$ in the RR's (Eqs. \ref{EqRRsA}), their solution for the polymerized phase can also be found exactly, being $R_2=R_4=R_5=R_7=R_8=0$, and $R_1=a/24 + (z + z^2/24)/a + z/24 - 1/2$, with $a \equiv \left( 36 z^2 + 216 z + z^3 + 24 \sqrt{3 z^3 + 81 z^2}\right) ^{1/3}$, $R_3=(1+2 R_1)/z$ and $R_6=\sqrt{R_1 (1+R_1)/z}$, in approaches A and B. In case C, one finds $R_1 = z-1/2$, $R_3=1$, $R_6=\sqrt{4 z-2}/2$ and $R_i=0$ otherwise. Noteworthy, this fixed point is independent of $\omega_1$, indicating that the thermodynamic properties of the model will not depend on this parameter, as expected.

In approaches A and B, the only non-null eigenvalue of the Jacobian matrix in NP phase is $\lambda=z$, so, this phase is stable for $z\leq 1$ and its spinodal is at $z=1$. The fixed point above for the P phase is physical (and stable) for $z \geq 1$ and its stability limit is at $z=1$. Therefore, the phase diagrams in both approaches present only a critical line at $z=1$ separating the NP and P phases, which is consistent with a $\theta$-point located at $z^{\theta}=1$ and $\omega_1^{\theta} \rightarrow \infty$, as already discussed. In case C, we have the same scenario, but the critical line is at $z=1/2$.

Notice that $\omega_2=0$ leads to $\rho_{NNN}=0$, so that all polymer chains are straight in cases A and B. As already noticed, in case C, the chains are not necessarily straight, since they can bend out the elementary squares and still have $\rho_{NNN}=0$. In all cases, $\rho_{NN}=0$ is also found, as expected, since the infinite repulsion between NNN monomers also forbids NN ones. The density of monomers $\rho$ is a monotonic increasing function of $z$, being $\rho= (2 z-1)/(4 z-1)$ in approach C. In the other approaches the expressions are too long to be given here, but one also finds $\rho=0$ at $z=1$ and $\rho \rightarrow 1/2$ for $z\rightarrow \infty$. On the square lattice this shall corresponds to a ``soft crystalline'' phase, where half of the rows (or columns) of the lattice are occupied by straight, parallel and alternating chains. Indeed, placing such aligned (repulsive) polymer chains on a square lattice is analogous to the athermal problem of place infinite rigid rods with NNN exclusion. We remark that athermal lattice gases with exclusion of neighbors have been considered in literature for several ranges of exclusion (or particle sizes) \cite{heitor,nath} as well as mixtures of them \cite{TiagoGRM}. Furthermore, isotropic-nematic transitions in rigid rods is a problem largely studied (see e. g. \cite{jurgenLC,jurgenLC2} for recent surveys). However, for the best of our knowledge, rigid rods with neighbor exclusion has been considered only in the case of dimers with NN exclusion \cite{dickman} only.

Anyhow, the case of infinite rods with NNN exclusion can be solved on the square lattice, for example, following the recent transfer matrix (TM) calculation by Stilck and Rajesh \cite{jurgenLC2}. Considering the limit of infinite rods, \textit{without} neighbor exclusion, on a roted square lattice yielding a diagonal TM, those authors showed that the (degenerated) spectrum of eigenvalues of the TM is given by $\Lambda_k=z^m$, with $m=0, 1, \ldots, L$ or $\Lambda_k=0$ \cite{jurgenLC2}. This result is for an infinite stripe (in vertical) with width $L$ and periodic boundary conditions in horizontal (for more details see \cite{jurgenLC2}). One notices that all rods are parallel in this limiting case and that an eigenvalue of type $z^m$ is associated to a state of the system with $m$ parallel rods in horizontal direction. So, it is easy to particularize these results for the case with infinite NNN repulsion, by noting that two rods can not occupy adjacent columns (or rows) of the lattice. This will simply reduce the number of states of the TM and, consequently, the spectrum of eigenvalues, which shall be $\Lambda_k=z^m$, with $m=0, 1, \ldots, L/2$ or $\Lambda_k=0$, since on a stripe of (even) width $L$ it is possible to exist at most $L/2$ parallel rods/chains, due to exclusion. Thence, the largest eigenvalue $\Lambda_l$ and, consequently, the free energy $f=\frac{1}{L}\ln \Lambda_l$ and the density of monomers $\rho = z\left( \frac{\partial f}{\partial z}\right) $ are: $\Lambda_l = 1$, $f=0$ and $\rho=0$, for $z\leq 1$; and $\Lambda_l = z^{L/2}$, $f=\frac{1}{2}\ln z$ and $\rho=1/2$, for $z>1$. Namely, the system undergoes a discontinuous transition at $z=1$ from an empty lattice (the NP phase - for $z\leq 1$) to a low density nematic phase (the ``soft crystalline'' polymer phase - for $z > 1$).

Although the Husimi solution yields the correct transition point $z=1$ (at least in the more realistic A and B approaches), the continuous transition found is more a feature of the hierarchical lattice. At first, it is unexpected to find a continuous transition in a mean-field calculation, while the real transition is discontinuous. One recall, notwithstanding, that similar inversions have been observed in other models for $\theta$-polymers, with discontinuous transitions observed in simulations and continuous ones in exact solutions on hierarchical lattices \cite{Tiago,Tiago2}.

\section{Final Discussions and conclusions}
\label{conclusions}

In summary, we have studied a generalized ISAW model - where different forces exist between nearest-neighbor (NN) and next-nearest-neighbor (NNN) monomers - on a Husimi lattice built with squares. Three definitions of second neighbors - or interactions between them - have been considered, which effectively overestimate, match or underestimate the number of NNN monomers, compared with square lattice. Since all approaches leads to analogous thermodynamic behaviors, this suggests that a similar scenario can exists also on the regular lattice.

\begin{figure}[b]
\begin{center}
\includegraphics[width=9.0cm]{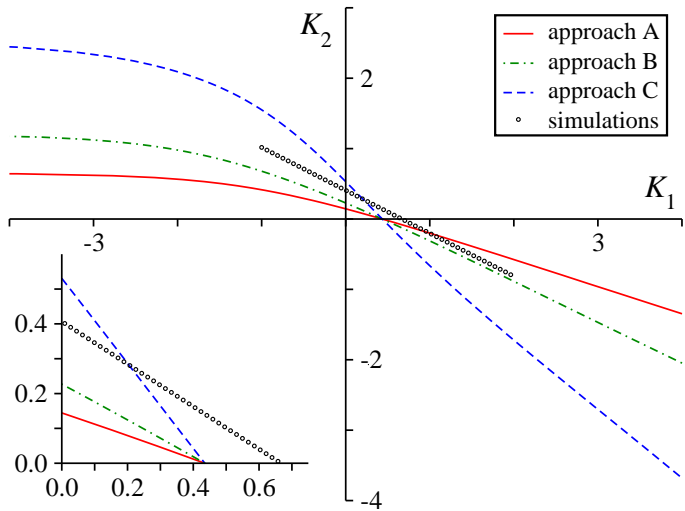}
\caption{(Color online) Canonical phase diagrams, in variables $K_2 = \ln \omega_2$ against $K_1 = \ln \omega_1$, for approaches A (full, blue) and B (dashed, red) and simulational (dash-dotted, black line) results from Ref. \cite{Nathann}. The inset shows the same data in the region of attractive interactions.}
\label{FigDiagCanonico}
\end{center}
\end{figure}

Indeed, our findings are in good agreement with the ones from Monte Carlo simulations on square and cubic lattices \cite{Nathann}, where only a $\theta$-line was observed, as well as with previous results from exact enumerations \cite{Lee2}. Interestingly, approximately linear $\theta$-lines were found  in the canonical phase diagrams reported in these works, around the region of positive energies. Figure \ref{FigDiagCanonico} shows the mapping of our grand-canonical phase diagrams in the canonical variables $K_2 \equiv \epsilon_2/k_b T (= \ln \omega_2)$ versus $K_1 \equiv \epsilon_1/k_b T (= \ln \omega_1)$ and, indeed, almost linear behaviors are found around the first quadrant of the diagrams, but the whole $\theta$-lines are curved. For comparison, the $\theta$-line found in simulations of the model on square lattice, $K_2 \simeq -0.6099 K_1 + 0.4066$ \cite{Nathann}, is also shown in Fig. \ref{FigDiagCanonico}. In the region corresponding to attractive interactions (highlighted in the inset), the $\theta$-line from approaches A and B are always below the one from simulations, which is expected, since mean-field results generally underestimate the (tri)critical points. A similar behavior is found for small $K_2$ in approach C, but as this parameter increases its $\theta$-line crosses the one from simulations, which is simply due to the underestimate in the number of NNN sites/monomers in this approach. Linear fits of the $\theta$-lines in the attractive region return the slopes $-0.333$ (in case A), $-0.522$ (in B) and $-1.221$ (in C), which are approximately 55\%, 86\% and 200\% of the value found in simulations. These behaviors are physically reasonable, because in a collapsing chain each monomer can have at most 2 NN monomers, while in approaches A, B and C it can have effectively a maximum of 6, 4 and 2 NNN monomers, respectively, which is consistent with $K_2 \sim -K_1/3$ (in A), $K_2 \sim -K_1/2$ (in B) and $K_2 \sim -K_1$ (in C).

\acknowledgments

This work was supported by CNPq and FAPEMIG (brazilian agencies). The author thanks JF Stilck for a critical reading of this manuscript and the kind hospitality of the group of Prof. James Evans at Iowa State University, where part of this work was done.

\appendix

\section{Free Energy}
\label{ApFreeEnergy}

The grand-canonical free energy of the model on the Cayley tree with $M$ generations is ${\tilde \Phi_M}=-k_B T \ln Y_M$ and one may, conveniently, define the adimensional free energy as $\Phi_M={\tilde \Phi_M}/k_B T$. Assuming that each surface site has a free energy $\phi_{s}$, while the ones in bulk has $\phi_{b}$ \cite{Gujrati}, it reads
\begin{equation}
 \Phi_M = N_{s}^M \phi_{s} + N_{b}^M \phi_{b},
\end{equation}
where $N_{s}^M$ and $N_{b}^M$ are the number of sites at surface and bulk, respectively, in generation $M$. Considering a Cayley tree built with squares and ramification $\sigma$ (coordination number $q=2(\sigma+1)$), these numbers are
\begin{equation}
N_{s}^M = 4 (3 \sigma)^{M-1} \quad \quad \text{and}  \quad  \quad  N_{b}^M = 4 \dfrac{(3 \sigma)^{M-1} -1}{3 \sigma -1}.
\end{equation}
From these equations, one finds
\begin{equation}
\phi_{b} = \dfrac{1}{4}\left[ \Phi_{M+1}- 3 \sigma \Phi_M \right] = -\dfrac{1}{4} \ln\left[\dfrac{Y_{M+1}}{Y_{M}^{3 \sigma}} \right],
\end{equation}
which is the reduced free energy in the bulk of the Cayley tree built with squares, i. e., the Husimi lattice.

As discussed in Sec. \ref{defmod}, in general, one may write $Y_M = (g_{0}^{M})^{4 \sigma} y$ and so $Y_{M+1} = (g_{0}^{M+1})^{4 \sigma} y$ (where $y=y_A$ or $y=y_B$ depending on the approach). In addition, it is easy to see that $g_{0}^{M+1} = (g_{0}^{M})^{3 \sigma} R_0$, then
\begin{equation}
\lim_{M \to \infty} \frac{Y_{M+1}}{Y_{M}^{3 \sigma}}=\frac{R_0^{4 \sigma}}{y^{3 \sigma - 1}},
\end{equation}
leading finally to
\begin{equation}
\phi_b=-\frac{1}{4}[ 4 \sigma \ln R_0 - (3 \sigma-1)\ln y].
\end{equation}

For the case $\sigma=1$, considered in this work, the expressions for $R_{0}$ are given in Eq. \ref{EqRRsA}h, while $y \equiv Y/g_{00}^{4}$ can be easily calculated from Eq. \ref{EqFuncPart1}, setting $\alpha=1$ (in approach A), $\alpha=1/2$ (in B) or $\alpha=0$ (in C).

\section{Tricritical lines}
\label{ApTricrical}

At the tricritical condition the solution of the recursion relations (RR's - Eqs. \ref{EqRRsA} and \ref{EqRRsB}) must be triply degenerated. Then, one may find the points at the parameter space where this happens, bearing in mind that they shall be on the NP spinodal.

In NP phase $R_{i}=1$ for $i=3$ and $R_{i}=0$ otherwise, so, near the critical surface (and the tricritical line), one may expand the RR's around, for example, $R_{6}$ keeping only the terms up to third order. A simple inspection of the RR's (Eq. \ref{EqRRsA}) shows that $R_i \simeq a_i R_{6}^2$ for $i=1,2,4,5$ and $8$, $R_3 \simeq 1 + a_3 R_{6}^2$ and $R_7 \simeq a_{7,1} R + a_{7,2} R_{6}^3$. Inserting this in the RR's and expanding them up to order $R_{6}^3$, one finds
\begin{subequations}
\begin{eqnarray}
\label{EqApendExpansionA}
D &\approx& 1 + C_0 R_{6}^2 \\
D a_1 R_6^2 &\approx& C_{1} R_{6}^2 \\
D a_2 R_6^2 &\approx& C_{2} R_{6}^2 \\
D (1+a_3 R_{6}^2) &\approx& 1 + C_{3} R_{6}^2 \\
D a_4 R_6^2 &\approx& C_{4} R_{6}^2 \\
D a_5 R_6^2 &\approx& C_{5} R_{6}^2 \\
D R_6 &\approx& C_{6,1} R_{6} + C_{6,2} R_{6}^3 \\
D (a_{7,1} R_6 + a_{7,2} R_{6}^3) &\approx& C_{7,1} R_{6} + C_{7,2} R_{6}^3 \\
D a_8 R_6^2 &\approx& C_{8} R_{6}^2 
\label{EqApendExpansionA2}
\end{eqnarray}
\end{subequations}
where
\begin{subequations}
\begin{eqnarray}
C_0 &=& 6 a_1+3 a_2+z a_8  \\
C_1 &=& z a_8+z^2 b^2  \\
C_2 &=& z^3 \omega_2 b^2  \\
C_3 &=& 2 (1+\omega_2^{\alpha}) a_1+2 \omega_2^{\alpha} a_2+2 a_1+a_2+z \omega_2 a_8  \\
C_4 &=& \omega_1 [z \omega_2^{\alpha} a_8+2 \omega_2 z^2 b (\omega_2^{\alpha}+\omega_2^{2\alpha} a_{7,1})]  \\
C_5 &=& 4 \omega_1^2 \omega_2^2 z^3 (\omega_2^{\alpha}+\omega_2^{2\alpha} a_{7,1})^2  \\
C_{6,1} &=& z b+\omega_2 z^2 b  \\
C_{6.2} &=& \omega_2 z^2 b (a_3+2 \omega_2^{\alpha} a_4+\omega_2^{2\alpha} a_5)+2 \omega_2^{\alpha+1} z^2 a_{7,2}+\omega_2 (2 \omega_2^{\alpha} a_1+\omega_2^{2\alpha} a_2) z^2 b \\  \nonumber
&+&2 \omega_1 z^2 \omega_2^{\alpha+1} a_8 (\omega_2^{\alpha}+\omega_2^{2\alpha} a_{7,1})+2 z \omega_2^{\alpha} a_{7,2}+[(2 (1+\omega_2^{\alpha})) a_1+2 \omega_2^{\alpha} a_2] z b  \\
C_{7,1} &=& 2 \omega_1 \omega_2^2 z^3 (\omega_2^{\alpha}+\omega_2^{2\alpha} a_{7,1})  \\
C_{7,2} &=& \omega_1 \omega_2 [\omega_2^{\alpha} z^2 a_8 b+2 \omega_2 z^3 (\omega_2^{\alpha}+\omega_2^{2\alpha} a_{7,1}) (2 a_3+4 \omega_2^{\alpha} a_4+2 \omega_2^{2\alpha} a_5)+2 \omega_2^{2\alpha+1} z^3 a_{7,2} \\  \nonumber
&+&\omega_1 \omega_2 (8 z^3 (\omega_2^{\alpha}+\omega_2^{2\alpha} a_{7,1})^3+2 z^3 \omega_2^{2\alpha} a_8 (\omega_2^{\alpha}+\omega_2^{2\alpha} a_{7,1}))]  \\  
C_8 &=& \omega_2 [z^2 b^2+8 \omega_1 z^3 \omega_2 (\omega_2^{\alpha}+\omega_2^{2\alpha} a_{7,1})^2]
\end{eqnarray}
\end{subequations}
with $b\equiv 1+\omega_2^{\alpha}+2 \omega_2^{\alpha} a_{7,1}$ and $\alpha=1$ (in case A), $\alpha=1/2$ (in B) or $\alpha=0$ (in C).

Equating the terms of same order in Eqs. (\ref{EqApendExpansionA})-(\ref{EqApendExpansionA2}), the relations $a_i = C_i$ for $i=1,2,4,5,8$ and $a_3 + C_0 = C_3$, $a_{7,1} = C_{7,1}$ and $a_{7,2}+a_{7,1} C_0 = C_{7,2}$ are obtained, allowing us to determine all $a_i$'s as functions of $z$, $\omega_1$ and $\omega_2$. Using these functions in the two additional equations $C_{6,1}=1$ - which leads to the same expression for the stability limit of the NP phase (Eq. \ref{EqNPlimitA}) - and $C_{6,2}=C_0$, the tricritical line is found. Although we do not find a closed expression for this line, it can be easily calculated with the help of an algebra software.

\end{document}